\documentclass[runningheads]{llncs}
\usepackage[T1]{fontenc}
\usepackage{graphicx}
\usepackage{xspace} 
\usepackage{tcolorbox}
\usepackage{listings}
\usepackage{color}
\usepackage{array}
\usepackage{colortbl}
\usepackage{hyperref}
\usepackage{multirow}
\usepackage{amsmath} 
\usepackage{amssymb}
\usepackage{orcidlink}
\usepackage{tabularx}

\newcolumntype{Y}{>{\centering\arraybackslash}X}

\hypersetup{
    colorlinks=true,
    linkcolor=blue,
    filecolor=magenta,      
    urlcolor=blue,
    pdftitle={CAISE25},
    pdfpagemode=FullScreen,
}

\usepackage{ulem} 
\newcommand{\noperations}{seven\xspace}

\newcommand{\serialization}{Pricing2Yaml\xspace} 
\newcommand{\model}{iPricings\xspace}

\definecolor{lightgray}{rgb}{.9,.9,.9}
\definecolor{darkgray}{rgb}{.4,.4,.4}
\definecolor{purple}{rgb}{0.65, 0.12, 0.82}
\definecolor{darkpurple}{rgb}{0.39, 0.10, 0.40}
\definecolor{backcolour}{rgb}{0.95,0.95,0.92}
\definecolor{codegray}{rgb}{0.5,0.5,0.5}
\definecolor{codegreen}{rgb}{0,0.6,0}

\lstdefinelanguage{Pricing4SaaS}{
  keywords={pricing, version, tiers, name, price, currency, description, feature, incuded},
  sensitive=true,
  keywordstyle=\color{darkpurple}\bfseries,
  ndkeywordstyle=\color{codegreen},
  identifierstyle=\color{black},
  comment=[l]{//},
  morecomment=[s]{/*}{*/},
  commentstyle=\color{codegreen}\ttfamily,
  stringstyle=\color{red}\ttfamily,
  morestring=[b]',
  morestring=[b]",
}

\lstdefinelanguage{IDL}{
  keywords={IF, THEN, AND, OR, NOT, Or, ZeroOrOne, OnlyOne, AllOrNone, LIKE},
  sensitive=true,
  keywordstyle=\color{darkpurple}\bfseries,
  ndkeywordstyle=\color{codegreen},
  identifierstyle=\color{black},
  comment=[l]{//},
  morecomment=[s]{/*}{*/},
  commentstyle=\color{codegreen}\ttfamily,
  stringstyle=\color{red}\ttfamily,
  morestring=[b]',
  morestring=[b]",
}

\lstdefinelanguage{Grammar}{
  keywords={Model:, Dependency:, RelationalDependency:, Param:, ArithmeticDependency:, Operation:, OperationContinuation:, ConditionalDependency:, Predicate:, Clause:, Term:, ParamValueRelation:, ClauseContinuation:, PredefinedDependency:, RelationalOperator:, ArithmeticOperator:, Not:, LogicalOperator:},
  sensitive=true,
  keywordstyle=\color{blue}\bfseries,
  ndkeywords={ID:, STRING:, DOUBLE:, PATTERN_STRING:, BOOLEAN:},
  ndkeywordstyle=\color{purple},
  identifierstyle=\color{black},
  comment=[l]{//},
  morecomment=[s]{/*}{*/},
  commentstyle=\color{codegreen}\ttfamily,
  stringstyle=\color{red}\ttfamily,
  morestring=[b]',
  morestring=[b]",
  alsodigit={:}
}

\lstdefinelanguage{MAP}{
  keywords={map, \{, \}, V, D, C},
  sensitive=true,
  keywordstyle=\color{black}\bfseries,
  ndkeywordstyle=\color{codegreen},
  identifierstyle=\color{black},
  comment=[l]{//},
  morecomment=[s]{/*}{*/},
  commentstyle=\color{codegreen}\ttfamily,
  stringstyle=\color{red}\ttfamily,
  morestring=[b]',
  morestring=[b]",
}

\lstdefinelanguage{YAML}{
  keywords={paths:, search:, get:, operationId:, parameters:, name:, x-dependencies:},
  sensitive=true,
  keywordstyle=\color{black},
  identifierstyle=\color{blue},
  comment=[l]{\#},
  commentstyle=\color{codegreen}\ttfamily,
  stringstyle=\color{red}\ttfamily,
  morestring=[b]',
  morestring=[b]",
  alsodigit={:, -}
}

\lstdefinestyle{listingtop}{
  float=tp,
  floatplacement=tbp,
  belowcaptionskip=-0.5cm,
}

\begin{document}
%
\title{Automated Analysis of Pricings in SaaS-based Information Systems}
%
\titlerunning{Automated Analysis of Pricings in SaaS-based Information Systems}
%
\author{Alejandro García-Fernández\orcidlink{0009-0000-0353-8891} \and
José Antonio Parejo\orcidlink{0000-0002-4708-4606} \and
Pablo Trinidad\orcidlink{0000-0002-1320-2424} \and
Antonio Ruiz-Cortés\orcidlink{0000-0001-9827-1834}}
\authorrunning{A. García-Fernández et al.}
%
\institute{SCORE Lab, I3US Institute, Universidad de Sevilla, Spain \\
\email{\{agarcia29,japarejo,ptrinidad,aruiz\}@us.es}}
\maketitle              

\begin{abstract}

Software as a Service (SaaS) pricing models, encompassing features, usage limits, plans, and add-ons, have grown exponentially in complexity, evolving from offering tens to thousands of configuration options. This rapid expansion poses significant challenges for the development and operation of SaaS-based Information Systems (IS), as manual management of such configurations becomes time-consuming, error-prone, and ultimately unsustainable. The emerging paradigm of Pricing-driven DevOps aims to address these issues by automating pricing management tasks, such as transforming human-oriented pricings into machine-oriented (iPricing) or finding the optimal subscription that matches the requirements of a certain user, ultimately reducing human intervention.
%
This paper advances the field by proposing \noperations analysis operations that partially or fully support these pricing management tasks, thus serving as a foundation for defining new, more specialized operations. To achieve this, we mapped iPricings into Constraint Satisfaction Optimization Problems (CSOP), an approach successfully used in similar domains, enabling us to implement and apply these operations to uncover latent, yet non-trivial insights from complex pricing models.
The proposed approach has been implemented in a reference framework using MiniZinc, and tested with over 150 pricing models, identifying errors in 35 pricings of the benchmark. Results demonstrate its effectiveness in identifying errors and its potential to streamline Pricing-driven DevOps.

\end{abstract}
\section{Introduction and Motivation}

Software as a Service (SaaS) has rapidly gained popularity in recent years \cite{Jiang2009}, driven by its flexibility to adapt to diverse user needs through subscription-based models. However, this adaptability has led to an exponential increase in configuration complexity, with modern pricing models evolving from offering a few options to thousands of potential configurations \cite{ICSOC2024}. For example, Salesforce’s November 2019 pricing offered up to 10 different configurations, while its July 2024 version allowed for over 12,000, as can be seen \href{https://sphere.score.us.es/pricings/sphere/Salesforce?collectionName=CAISE%202025}{here}. This rapid expansion presents significant challenges for the development and operation of SaaS-based Information Systems (IS), as manual management of these configurations becomes a demanding task ---being time-consuming and error-prone--- especially when adapting to frequent pricing changes.

Addressing these challenges requires a shift towards automated, scalable solutions. The paradigm of \emph{Pricing-driven Development and Operation} has emerged to tackle this issue, aiming to automate tasks related to the management of pricing models\footnote{For brevity, the term ``pricing models'' and ``pricings'' will be used interchangeably.} in SaaS-based Information Systems \cite{ICSOC2024}. These may include, for example, designing pricing models, converting human-oriented representations of pricings into machine-oriented (iPricings), or finding the optimal subscription that meets particular user requirements. 

A foundational step toward this vision was presented in CAISE Forum, where a metamodel for representing pricings was proposed \cite{CAISEFORUM2024}, and has laid the groundwork for the idea of machine-oriented pricings (iPricings) \cite{AI4PRICING}. Building on this foundation, subsequent efforts have focused on leveraging iPricings to develop solutions that automate tasks within the scope of Pricing-driven DevOps. For instance, \cite{ICWE2024} proposed a reference architecture for automating the variability management in client-server architectures leveraging feature toggling \cite{FOWLER2023}, and AI4Pricing introduced initial advancements towards automating the transformation between human-oriented pricings and iPricings \cite{AI4PRICING}. Although these contributions represent early steps, they highlight the progress being made toward realizing the vision of Pricing-driven DevOps. 

\begin{figure}
    \centering
    \frame{\includegraphics[width=\linewidth]{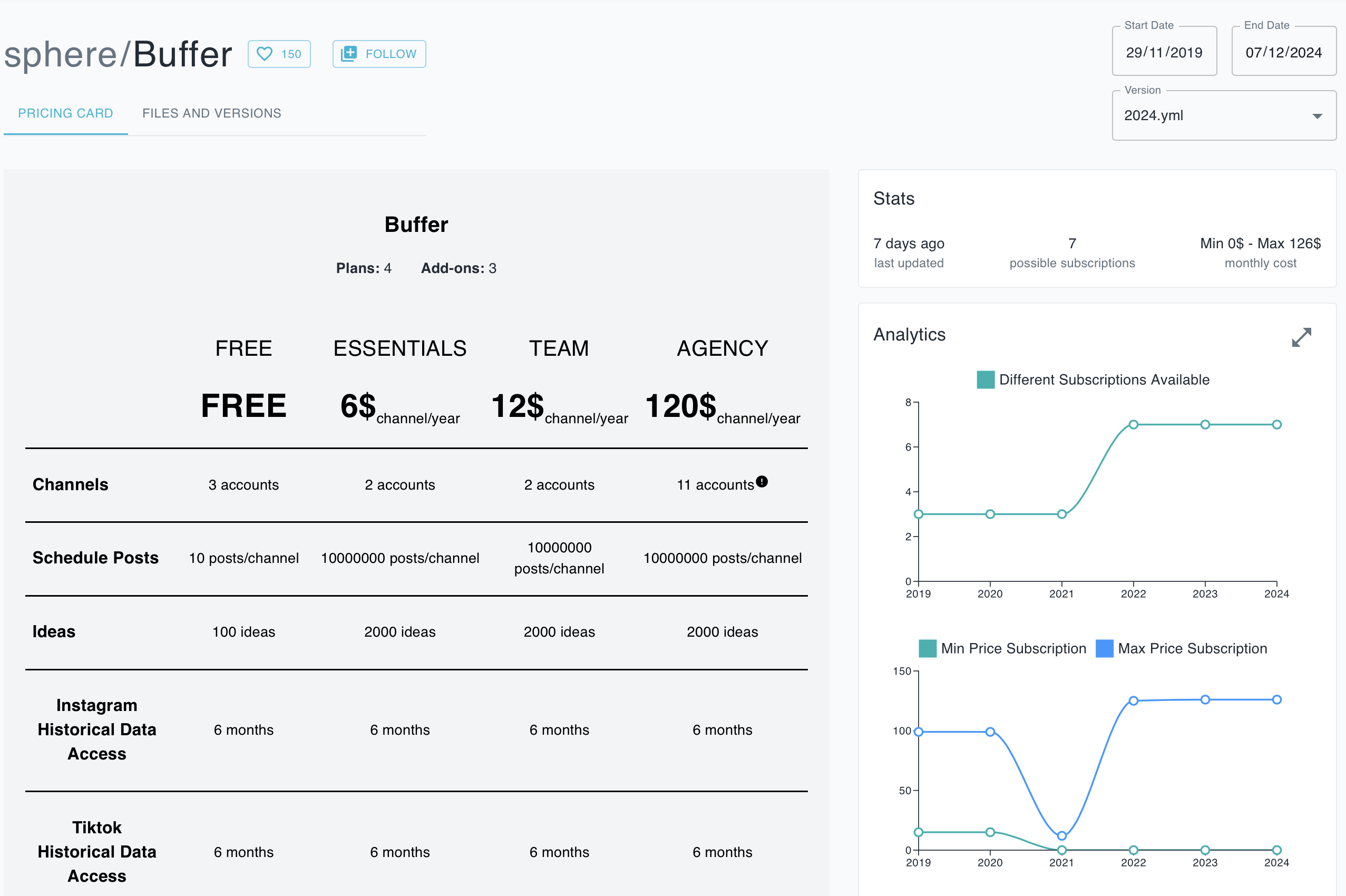}}
    \caption{A sample final product integrating the proposed set of analysis operations. Public access available \href{https://sphere.score.us.es/pricings/sphere/Buffer?collectionName=CAISE\%202025}{here}, along with other examples}
    \label{fig:pricingCard}
\end{figure}

However, a crucial pillar remains unexplored: the automated analysis (AA) of iPricings, i.e. the process of uncovering latent, yet non-trivial, information from machine-oriented pricings. For instance, this approach has the potential to enable real-time validation of pricing structures, strengthen data-driven pricing decision-making, or assist customers in identifying the most cost-effective subscription that meets their specific requirements. In light of this, the paper introduces the following original contributions, inspired by solutions previously applied in the automated analysis for feature models in CAISE 2005 \cite{Caise2005} and service level agreements \cite{muller2013automated}:

\begin{enumerate}

    \item An unprecedented formalization of iPricings as Constraint Satisfaction Optimization Problems (CSOPs), enabling their automated analysis through constraint programming techniques.
    
    \item A set of \noperations fundamental analysis operations, built on the CSOP-based formalization, to support pricing structural validation, configuration space cardinality computation, and optimal subscription selection based on user requirements.

    \item A tool to apply these operations, built with the MiniZinc solver \cite{minizinc} and integrated within a real platform (see Fig. \ref{fig:pricingCard}).

    \item Two curated datasets: 

    \begin{itemize}

        \item A novel synthetic dataset comprising 20 pricing models with seeded consistency errors. Although our solution successfully detected most of them, it missed 6 instances, making this dataset a valuable resource for future work aimed at improving automated inconsistency detection.
    
        \item An updated version of the benchmark introduced in \cite{ICSOC2024}, incorporating fixes for the 35 structural issues identified by our approach, providing a more reliable baseline for future research.
    \end{itemize}
\end{enumerate}

The remainder of this paper is structured as follows: Section \ref{sec:background} frames the contribution within the existing literature and presents the core concepts, challenges, and research advancements in SaaS pricing. Section \ref{sec:analysis} presents an automated approach for iPricings analysis using CSOP-based operations. Next, Section \ref{sec:tooling-support-and-validation} presents a reference implementation of the approach and tests it using benchmark and synthetic datasets. Finally, Section \ref{sec:conclusions} showcases the conclusions.

\section{Background \& Related Work}
\label{sec:background}
\subsection{User-oriented SaaS Pricings}

A pricing is a part of a SaaS customer agreement \cite{garcia2021flexible}. It structures the \textit{features} of a service ---defined as the distinctive characteristics whose presence/absence may guide a user’s decision towards a particular subscription \cite{CAISEFORUM2024}--- into \textit{plans} and \textit{add-ons} to control users' access to such features. While users can only subscribe to one of the available plans, they can subscribe to as many add-ons as they want, as long as they are available for the contracted plan. As can be noted, in this domain, a feature encompasses both \textit{functional features}, which constitute the core software product, and \textit{extra-functional features}, which, while external to the product itself, enhance its perceived value (e.g., support, SLA guarantees, etc). By including both types of features, pricings capture a more comprehensive view of what influences customer value, aligning the service offering with the overall customer agreement. 

\begin{figure}[htb]
    \centering
    \includegraphics[width=\textwidth]{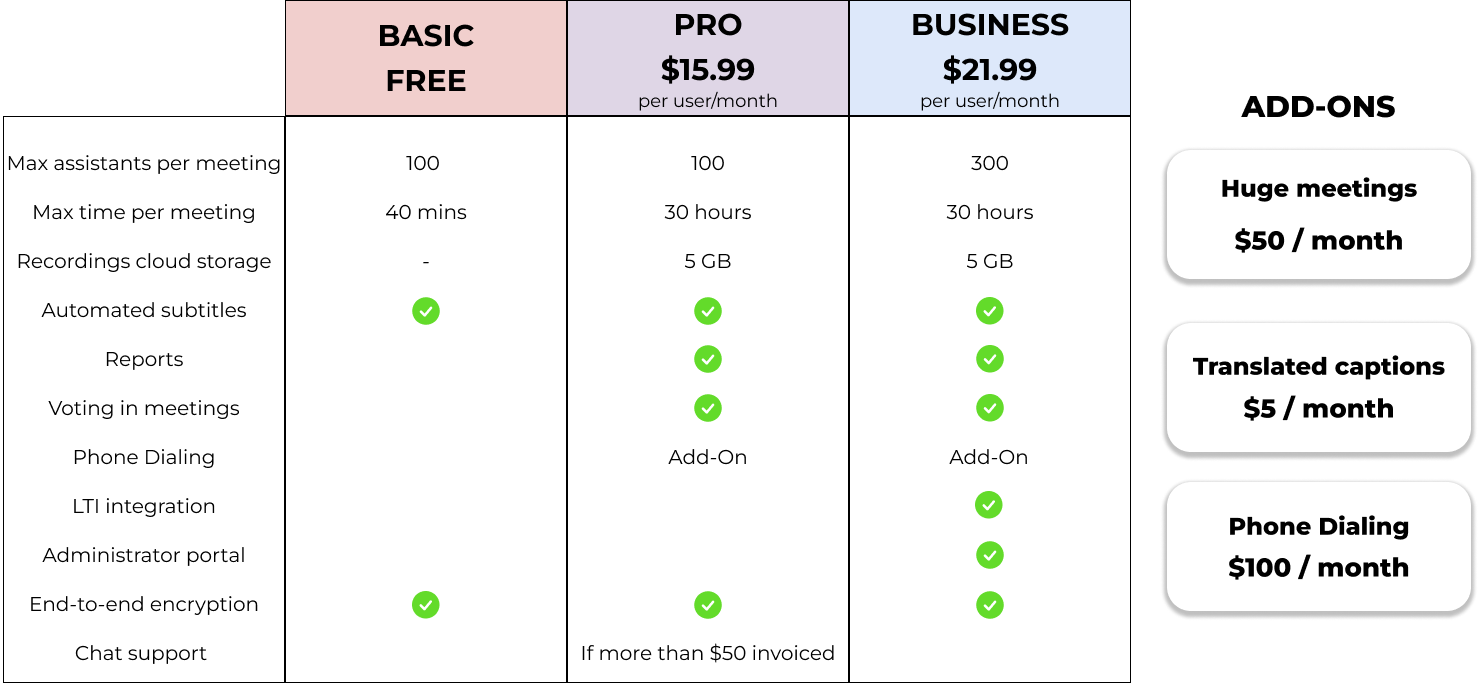}
    \caption{Excerpt of Zoom’s pricing with 13 features, three plans, and three add-ons, with a configuration space size of 20}
    \label{fig:zoomPricing}
\end{figure}

In order to illustrate the upcoming concepts, we will use a real-world running example: \href{https://zoom.us/pricing}{Zoom}. This is a cloud-based video conferencing service that enables users to virtually meet with others and optionally record the sessions to watch them later. An excerpt of its pricing, consisting of 13 features, is presented in Fig. \ref{fig:zoomPricing}.\footnote{Pricing entries that impose or extend limits on meetings are considered usage limits rather than individual features—the overarching feature in this case is “meetings”.} In this example, nine features are managed through plans, one feature is associated with an add-on (``translated captions''), and one is governed by both (``phone dialing''). The pricing also enforces usage limits on the ``meetings'' feature (e.g., maximum assistants per meeting and maximum meeting duration) meaning that although the feature is available in all plans, the extent of their usage differs ---higher-priced plans offer higher limits.


\subsection{Machine-oriented SaaS Pricings (iPricings)}

\begin{figure}
    \centering
    \includegraphics[width=\linewidth]{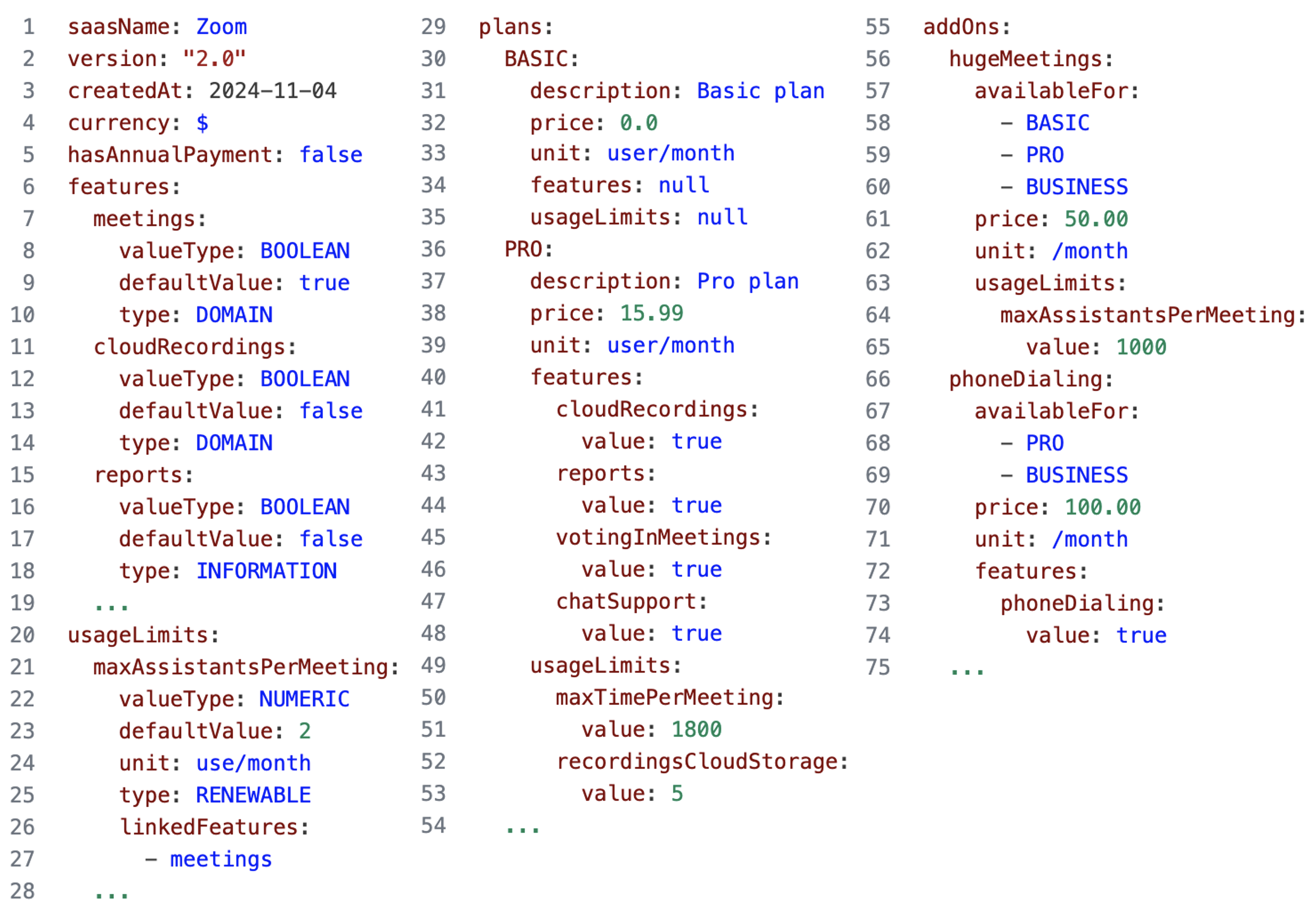}
    \caption{Zoom's pricing excerpt serialized in Pricing2Yaml}
    \label{fig:pricing2yaml}
\end{figure}

The paradigm of \emph{Pricing-driven Development and Operation} has emerged to address the challenges associated with pricing management tasks. Despite growing interest, there has been limited progress in this area. The Pricing4SaaS model, introduced as a formal metamodel, represents one of the first generalized attempts to standardize pricings \cite{CAISEFORUM2024}. This model defines pricings as a combination of \textit{plans} and \textit{add-ons} that regulate access to a set of features whose usage may be limited by usage limits. Additionally, its YAML-based serialization, Pricing2Yaml (see Fig. \ref{fig:pricing2yaml}), has served as the initial step towards automated tooling for Pricing-driven DevOps \cite{ICWE2024} and, therefore, iPricings.

Building on this foundation, the study undertaken by \cite{ICSOC2024} represents the first major effort to systematically analyze SaaS pricings. By modeling up to 162 pricings from 30 different SaaS providers –up to 6 different pricing versions between 2019 and 2024 per SaaS– this study not only demonstrated the flexibility of \serialization (and, by extension, the adaptability of Pricing4SaaS) but also revealed significant trends in the evolution of pricings.

To perform their analysis and measure such evolution, the authors introduced the concept of  \textit{configuration space,} i.e. the set of different subscriptions within a pricing. As an illustration, the size of the configuration space of Zoom's pricing excerpt (see Fig. \ref{fig:zoomPricing}) is 20. Their results revealed an exponential growth of the configuration space, primarily due to the addition of add-ons over the last years, which diversifies available subscriptions and increases customization. Moreover, the study also highlights a widening gap between this configurability and the actual capabilities of feature toggling tools, which often struggle to manage the increasing variability. This disparity indicates that existing technical support for handling such challenges remains limited, underscoring a need for automated tools that reduce complexity and time-to-market while allowing the software to adapt to evolving customer demands.

\subsection{Automated Analysis in Feature Models and SLAs}
\label{sec:related-work}

Automated pricing analysis shares foundational elements with existing paradigms like Feature Models (FM) in Software Product Lines (SPLs) \cite{benavides10} and SLA analysis frameworks such as WS-Agreement \cite{muller2013automated}. Both FMs and pricings leverage features as their minimal unit of representation, functioning as variability modeling languages (VMLs) \cite{Caise2005} that enable configurability and modularity. However, pricings extend beyond the usually technical scope of FMs by incorporating operational and market-driven elements critical to SaaS offerings, such as SLA guarantees, support, discounts, etc. Although most of those elements could be represented through attributed or stateful feature models \cite{Trinidad2013}, using DSLs and tools with a syntax and operations more specific to the pricings management problem emerges as a promising, more natural approach. Moreover,  pricings must handle dynamic variables, such as usage-based limits, which can vary depending on add-ons (e.g., overage fees) or be customized by the user when subscribing, making their representation as a FM more complex.



Similarly, the automated analysis of iPricings shares conceptual ground with WS-Agreement, a framework for formalizing service agreements through an XML-based serialization, primarily focusing on SLAs \cite{Muller13TSC}. While WS-Agreement supports the evaluation of user-defined constraints (e.g., agreement creation constraints), it lacks the vocabulary to represent pricing-specific elements like plans, add-ons, and usage-based fees. Extending WS-Agreement to encompass iPricings would require substantial semantic modifications, making it verbose and poorly suited to SaaS pricing strategies. Moreover, in contrast, iPricings approaches such as \cite{CAISEFORUM2024,ICSOC2024} embed subscription constraints directly within the pricing artifact (e.g., “select one plan and up to three add-ons” or “if add-on A2 is selected, exclude A7”), enabling a seamless and efficient evaluation of customer configurations without requiring external frameworks or additional layers of complexity.

The automated analysis of iPricings can thus be viewed as a specialized subset of SLA analysis, where the SLA models have variants corresponding with each possible configuration users can subscribe to. In this way, the demonstrated efficacy of CSOPs in the automated analysis of WS-Agreement documents for evaluating SLA compliance \cite{Muller13TSC,Muller18TSC}, combined with the conceptual parallels between WS-Agreement and iPricings, has been the driving force behind this work’s adoption of CSOPs as its approach to the AA of iPricings. With this alignment, we ensure a structured methodology that leverages proven techniques to address the unique challenges posed by iPricings. Pricing4SaaS builds upon this foundation by categorizing SLA-related constraints, such as guarantees (e.g., TTO) and support, and embedding them within pricings as features and usage limits. 

This duality positions iPricings as the intersection of two paradigms: the configurability and modularity inherent to feature models, and the guarantee-driven focus of WS-Agreement. By combining these dimensions, iPricings not only address the operational intricacies and market-driven complexities of SaaS offerings but also establish themselves as a versatile variability modeling language capable of capturing both functional and service-level aspects of modern software systems.

\section{Automated Analysis of \model}
\label{sec:analysis}

Given the growing complexity of pricings and their fast pace of change, automating the analysis of these structures is essential. Automation enables real-time validation of pricing specifications, ensuring their consistency and providing critical insights for tasks such as design, management of changes, and the choice of specific subscriptions based on additional user requirements.

In what follows, we present our approach for the automated analysis of \serialization specifications (see Fig. \ref{fig:pricing2yaml}) using constraint programming, given that, to the best of our knowledge, is the only serialization of pricings proposed in the literature. Along this section, we first present the formal semantics of \serialization by explaining how these specifications can be mapped to a constraint satisfaction optimization problem (CSOP). Then, we present a catalog of \noperations analysis operations of \serialization specifications and show how they can be automated using standard constraint programming reasoning operations.

\subsection{Formal Semantics of \model}
\label{sec-idlSemantics}

The primary objective of formalizing \model is to establish a sound basis for automating the analysis, decision-making, and task automation in the context of SaaS pricing models, thus enacting the vision of iPricings. Following the principles defined by \cite{Hofstede98}, we follow a transformational style by translating \serialization specifications to a target domain suitable for the automated analysis (\emph{Primary Goal Principle}). Specifically, we propose Constraint Satisfaction Problems (CSPs) as this target domain, enabling pricings to be analyzed using constraint programming tools. This approach aligns with previous efforts to automate the analysis of feature models \cite{benavides10} and service level agreements \cite{Muller13TSC,Muller18TSC}.

A CSP is a 3-tuple $(V, D, C)$, where $V$ is a set of variables, $D$ represents the set of domains for these variables, and $C$ is a set of constraints among the variables. A solution to a CSP is an assignment of values from each domain in $D$ to the variables in $V$ that satisfies all constraints in $C$. Solving a CSP is finding one or more solutions that meet all the  constraints.  Constraint Satisfaction Optimization Problems (CSOPs) extend the CSP formulation by incorporating an objective function. Formally, a CSOP can be represented as $(V, D, C, O)$, where $(V, D, C)$ defines the underlying CSP, and $O$ is an objective function that assigns a value (often to be minimized or maximized) to each possible solution. Thus, instead of merely satisfying the given constraints, a CSOP seeks a solution that also optimizes $O$.

Our proposal formulates pricings as CSOPs. We introduce a generic model $\pi$ that captures the common structure and constraints of all pricing scenarios, as well as a mapping from Pricing2Yaml to a data file specifying the parameters of a given instance. This clear separation between the model ---defining the structure of the pricing problem--- and the data ---specifying a particular instance--- enables the creation of a single, reusable model for multiple datasets, simplifying the formulation, analysis, and optimization of pricings as CSOPs.

\begin{figure}
    \centering \includegraphics[width=0.8\linewidth]{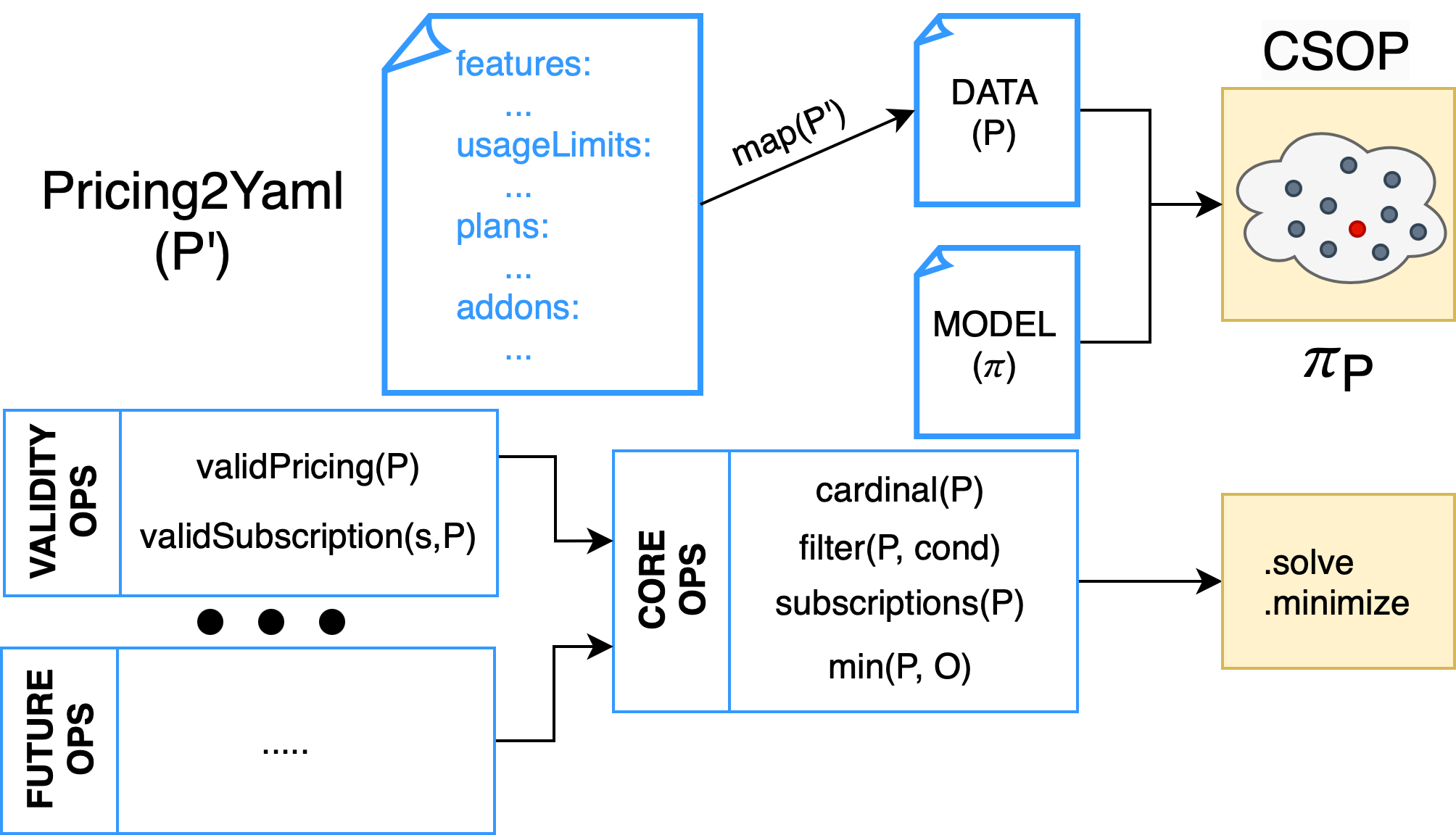}
    \caption{Outline of our approach to automate the analysis of \serialization serializations}
    \label{fig:approach}
\end{figure}

In our approach (see \ref{fig:approach}), the model $\pi$ represents a generic pricing scenario, with $P_p$ plans and $P_A$ add-ons. It encapsulates fundamental validity criteria for pricings that stem directly from the semantics of the pricing concepts (plans, add-ons, etc.). More specifically, $\pi$ enforces the following constraints:

\begin{enumerate} 

    \item A pricing $P$ cannot be empty, it must at least contain a plan or an add-on

    $$
        notEmpty(P) \iff (P_p \neq \emptyset \lor P_A \neq \emptyset)
    $$



    \item For each plan $p \in P_p$, if usage limits are defined, the features affected by those usage limits must be included within the same plan.

    $$
        containsLinkedFeatures(p) \iff 
        \begin{cases}
            p.usageLimits \neq \emptyset \implies \\
            \forall u \in p.usageLimits, \\
            u.linkedFeatures \subseteq p.features
        \end{cases}
    $$

    \item If the pricing includes both plans and add-ons, each add-on $a \in P_A$ must be available for at least one plan.

    $$
        isAvailable(a) \iff ((P_p \neq \emptyset \land P_A \neq \emptyset) \implies a.availableFor \neq \emptyset)
    $$
\end{enumerate}
In a very similar way, the validity criteria for a subscription $S$ comprises the following constraints:
\begin{enumerate}

    \item A subscription cannot be empty; it must include either a plan $p$ with zero or more add-ons (the set of selected add-ons is denoted as $A_S$) if the pricing includes plans, or at least one add-on if the pricing consists solely of add-ons.

    $$
        notEmptySubscription(p, A_S) \iff (p \neq \emptyset \lor A_S \neq \emptyset)
    $$

    \item If the pricing includes both plans and add-ons, and a subscription includes add-ons, every add-on must be available for the specific plan selected in that subscription.

    $$
        addonsAvailable(p, A_S) \iff \begin{cases}
            p \neq \emptyset \land A_S \neq \emptyset \implies \\
            \forall a \in A_S, \\
            p \in a.availableFor
        \end{cases}
    $$
    \item If a subscription includes add-ons with dependencies on other add-ons, all required dependencies must also be part of the subscription.

    $$
        dependencyAware(A_S) \iff \begin{cases}
            A_S \neq \emptyset \implies \\
            \forall a \in A_S, \forall a' \in a.dependsOn, \\
             a' \in A_S
        \end{cases}
    $$

    \item If a subscription includes add-ons that exclude others, all excluded add-ons cannot be included within the subscription.

    $$
        exclusionAware(A_S) \iff \begin{cases}
            A_S \neq \emptyset \implies \\
            \forall a \in A_S, \forall a' \in a.excludes, \\
             a' \not\in A_S
        \end{cases}
    $$
\end{enumerate}



Then, we denote every model instance as $\pi_P$, created by pairing the model $\pi$ with the particular data of a pricing $P$. Solutions to $\pi_P$ ---i.e. $solve(\pi_{P})$--- thus represent the set of feasible configurations, for a pricing $P$, where all defined constraints are satisfied.


\subsection{Analysis Operations}
\label{sec-analysisOperations}

Building on the formalization outlined in the previous section, where iPricings were mapped to CSOPs, we now turn our attention to the analytical operations that this formalization enables. These operations are key to extracting meaningful, latent insights from iPricing specifications. To provide clarity and continuity, each operation is illustrated using the Zoom’s pricing excerpt (see Fig. \ref{fig:zoomPricing}) as a running example.

\subsubsection{Number of Configurations}

One of the most important metrics derived from a pricing is the cardinality of its configuration space, representing the number of different configurations that can be generated from it. A higher cardinality indicates increased flexibility for the SaaS, but also adds complexity to its maintenance.

\begin{definition}[Cardinality] Let P be a pricing, the number of configurations of P, hereinafter cardinal, is equal to the number of solutions of its equivalent CSOP $\pi_P$.
\end{definition}

\vspace{-0.6cm}

$$
cardinal(P) = |solve(\pi_P)|
$$

In the example of Zoom's pricing excerpt, $cardinal(Zoom) = 20$. However, simply by adding a new plan, the cardinality increases to 28. In contrast, introducing an additional add-on doubles the cardinality, resulting in $cardinal(Zoom) = 40$. As demonstrated in \cite{ICSOC2024}, each new add-on leads to an exponential increase in cardinality, whereas adding a new plan results in only a linear increase.

\subsubsection{Filter}

This is particularly useful for customers that are looking for a configuration with specific features and usage limits, i.e. they are not interested in all possible configurations but those that tailor to their needs ---i.e. pass the filter.

\begin{definition}[Filter]
    Let $P$ be a pricing and $F$ a constraint representing a filter, containing the desired features and usage limits' values within the subscription. The filtered pricing of $\pi_P$, hereinafter filter, is equal to $\pi_P$ with the added constraint $F$.
\end{definition}

$$
filter(P, F) \iff (\pi_P \land F)
$$

A possible filter for zoom's pricing excerpt would be to ask for all configurations with the ``administrator portal'' feature and, at least, a 200 assistants capacity in meetings. By applying this filter to the pricing, the number of potential configurations decreases from 20 to 8.
$$
F = (administratorPortal = true \land maxAssistantsPerMeeting >= 200)
$$
$$
cardinal(filter(Zoom,F)) = 8
$$

\subsubsection{Subscriptions}

Once $\pi_P$ is defined, there should be a way to get all the solutions of the CSOP, i.e. possible configurations of the pricing.

\begin{definition}[Subscriptions]
    Let P be a pricing, the potential configurations of the pricing, hereinafter subscriptions, are equal to the solutions of the equivalent CSOP $\pi_P$.
\end{definition}

$$
subscriptions(P) = \{s \in solve(\pi_P)\}
$$

Within zoom, we would like to get all possible subscriptions that include the ``records'' feature. In this way, $P = filter(Zoom, records = true)$ and $subscriptions(P) = \{s \in solve(\pi_P \land records = true)\}$.

\subsubsection{Subscription Cost}

In order to perform cost optimization operations, it is necessary to define how the cost of a subscription is computed.

\begin{definition}[Subscription Cost]
    Let $s$ be a subscription of the pricing $P$. The cost associated with $s$ is calculated as the sum of the selected plan's price, if it exists, and the prices of the selected add-ons, if any.
\end{definition}

$$
cost(s) = s.plan.price + \sum_{a \in s.addons} a.price 
$$


Within the zoom's pricing excerpt, we would like to get the cost of a subscription that includes the plan ``PRO'', and the add-on ``Huge Meetings'', i.e., $s = (PRO, \{Huge Meetings\})$. Thus, $cost(s) = 65.99$.

\subsubsection{Pricing Validation}

A pricing is valid when it has at least one configuration that can be selected. That is, a model where $\pi_P$ has at least one solution.

\begin{definition}[Valid Pricing]
    A pricing $P$ is valid if its equivalent CSOP $\pi_P$ is satisfiable.
\end{definition}
\vspace{-0.3cm}
$$
valid(P) \iff subscriptions(P) \neq \emptyset
$$

As an illustration, let $CircularConstraints$ be a pricing that consists solely of three add-ons $P_A = \{a_1, a_2, a_3\}$, that share the following restrictions among them: i) $a_1$ depends on $a_2$, ii) $a_2$ depends on $a_3$, and iii) $a_3$ excludes $a_1$. Then:

$$
valid(CircularConstraints) = false
$$


\subsubsection{Subscription Validation}

Since many different subscriptions can be generated from a pricing, it should be possible to contrast whether a subscription is valid according to a pricing.

\begin{definition}[Valid Subscription]
    Let $P$ be a pricing and $s$ be a subscription, that is considered valid if and only if it is a potential solution of $\pi_P$.
\end{definition}
\vspace{-0.1cm}
$$
valid(s,P) \iff s \in subscriptions(P)
$$

Taking the excerpt of zoom's pricing as an illustration, a subscription that includes the usage limit ``maxAssistantsPerMeeting = 1200'' is not valid, as the maximum permitted value for this usage limit is 1000, achievable only by purchasing the ``Huge Meetings'' add-on.

\subsubsection{Optimum Subscription}

Identifying the optimum subscription based on a specific criterion is essential in pricing for both customers and providers. Customers can determine the minimum price for a defined set of features and usage limits, while providers gain insight into the price range at which a particular set of features and usage limits is offered.

\begin{definition}[Optimum]
    Let P be pricing and O an objective function for the subscription cost, then the optimum set of subscriptions, hereinafter max and min, is equal to the optimum space of $\pi_P$.
\end{definition}

$$
max(P,O) = max(\pi_P, O)
$$
$$
min(P,O) = min(\pi_P, O)
$$

It is also possible to apply a filter to the zoom's pricing excerpt and then ask for an optimal subscription. Thus, a possible optimum criterion for our running example would be to ask for all subscriptions with $record = true \land cloudStorage >= 5$, and the minimum value for the subscription cost. In this case, optimum subscriptions $S_{opt}$ are:

$$
S_{opt} = min(filter(Zoom, record = true \land cloudStorage >= 5), cost)
$$

\section{Tooling Support and Testing}
\label{sec:tooling-support-and-validation}

This section presents the implementation of the proposed CSOP-based approach for automating iPricing analysis using MiniZinc \cite{minizinc}, as well as its testing with real-world and synthetic datasets. Due to space limitations, the specific mapping of each individual constraint from Section \ref{sec-analysisOperations} to this solver, as well as the definition of the involved variables and domains, are defined in the technical report of this paper, that is available as part of the laboratory package~\cite{LABPACK}.

At the core of this implementation is MiniZinc, a medium-level declarative modeling language recognized for its flexibility and compatibility with a wide range of constraint programming (CP) solvers. By separating CSPs into two components ---the model, which defines the constraints and structure of a class of problems--- and the data ---which specifies the parameters for a particular instance--- MiniZinc allows a single model to be reused across datasets, thus enhancing scalability and modularity. Moreover, its ability to translate CSPs into the FlatZinc format ensures solver compatibility \cite{minizinc}, bridging the gap between high-level problem descriptions and low-level solver implementations.

Leveraging MiniZinc’s modularity, we developed a hierarchical formalization for iPricings, structured using a bottom-up methodology.
We begin with a foundational model that establishes the core parameters and variables of the CSOP (\href{https://github.com/isa-group/SaaS-analysis/tree/CAISE'25-Research-Track-v2/data/models/minizinc/PricingModel.mzn}{PricingModel.mzn}). The parameters, drawn from input data, encapsulate the details of a specific pricing, while the variables define the subscription attributes that constitute the solutions of the CSOP. This foundation enables us to incrementally add layers of constraints that are tailored to specific analytical tasks, making it easier to extend the set of AA operations the approach can support.

Building on this foundation, we introduce models that ensure the integrity of the input pricing (\href{https://github.com/isa-group/SaaS-analysis/tree/CAISE'25-Research-Track v2/data/models/minizinc/operations/validation/valid-pricing.mzn}{valid-pricing.mzn}) and the correctness of derived subscriptions (\href{https://github.com/isa-group/SaaS-analysis/tree/CAISE'25-Research-Track-v2/data/models/minizinc/operations/validation/valid-subscription.mzn}{valid-subscription.mzn}). On top of these, we built specialized analysis models to perform the operations described in Section \ref{sec:analysis}. Each layer refines and extends the constraints set by its predecessor, ultimately producing a fully tailored CSOP for a specific task. To instantiate an analysis, the chosen model is paired with a corresponding \textit{.dzn} data file ---automatically generated from a Pricing2Yaml serialization--- ensuring a clear separation between the generic problem structure and the pricing-specific details.

To validate the proposed framework, two datasets were utilized. The first one, derived from the benchmark presented in \cite{ICSOC2024}, encompasses 162 real-world SaaS pricings spanning 30 different services over a five-year period (see Tab.~\ref{table:dataset2024}). The initial set of services was taken from the repository presented in \cite{CAISEFORUM2024}, and subsequently expanded based on three main criteria: (i) snapshot availability via the \href{https://web.archive.org}{Wayback Machine}, (ii) clear feature listings within each snapshot's pricing page, and (iii) appropriate temporal spacing between snapshots (6 to 12 months). Although the resulting dataset captures services of various sizes and key real-world aspects such as frequent pricing updates and usage-based offerings, the exclusive use of Pricing2Yaml serialization causes inherent limitations in representing more complex scenarios, such as bundled offerings or discounts.

Analysis of this dataset revealed inconsistencies in 35 out of 162 pricings (21.6\% of the cases): i) 25 linked feature mismatches, where non-zero usage limits were assigned to unavailable features; ii) eight incorrect feature definitions, where features were incorrectly represented as numerical values instead of usage limits; iii) one temporal error, with a pricing having a future creation date; and iv) one pricing without specific prices (\href{https://web.archive.org/web/20201101013956/https://trustmary.com/pricing/}{Trustmary 2020}), where all plans required contacting sales. Notably, 34 out of 35 inconsistencies stemmed from errors during the manual transition from web pricings to iPricings, underscoring the risk of human error in this process. To address this, we corrected all 34 inconsistencies originating from human errors, ensuring the dataset aligns with the intended specifications.\footnote{The corrected version of the real-world pricings dataset is included both in the laboratory package~\cite{LABPACK} and in \href{https://sphere.score.us.es/pricings}{SPHERE}.} Additionally, to mitigate similar issues in the future, we have developed a Pricing2Yaml editor within SPHERE, a platform for intelligent pricing-driven solutions.\footnote{\url{https://sphere.score.us.es/editor}} The editor provides real-time error detection during the modeling process (using the validity constraints defined in this paper) and offers a live rendering of the pricing as it is constructed/edited, streamlining the development of reliable iPricings.

\begin{table}[htbp]
\centering
\begin{tabularx}{\textwidth}{|l|Y|Y|Y|Y|c|l|Y|Y|Y|Y|Y|}
\hline
\textbf{SaaS} &
\textbf{S} & \textbf{F} & \textbf{P} & \textbf{A} & \textbf{C} &
\textbf{SaaS} &
\textbf{S} & \textbf{F} & \textbf{P} & \textbf{A} & \textbf{C} \\ \hline
\rowcolor{gray!20}
\href{https://www.salesforce.com/eu/sales/pricing/}{Salesforce} & 6 & 111 & 3 & 14 & 12544 &
\href{https://buffer.com/pricing}{Buffer} & 6 & 76 & 4 & 3 & 7 \\ \hline
\href{https://github.com/pricing}{GitHub} & 6 & 81 & 3 & 14 & 8960 &
\href{https://www.atlassian.com/software/jira/pricing}{Jira} & 6 & 60 & 4 & 1 & 7 \\ \hline
\rowcolor{gray!20}
\href{https://www.postman.com/pricing/}{Postman} & 5 & 100 & 4 & 12 & 1412 &
\href{https://www.notion.so/pricing}{Notion} & 4 & 58 & 4 & 1 & 7 \\ \hline
\href{https://databox.com/pricing}{Databox} & 6 & 62 & 5 & 8 & 786 &
\href{https://www.figma.com/pricing/}{Figma} & 6 & 90 & 6 & 0 & 6 \\ \hline
\rowcolor{gray!20}
\href{https://www.openphone.com/pricing}{OpenPhone} & 5 & 48 & 3 & 6 & 192 &
\href{https://www.box.com/pricing}{Box} & 6 & 50 & 5 & 0 & 5 \\ \hline
\href{https://www.wrike.com/comparison-table/}{Wrike} & 6 & 78 & 5 & 5 & 85 &
\href{https://www.canva.com/pricing/}{Canva} & 6 & 92 & 4 & 0 & 4 \\ \hline
\rowcolor{gray!20}
\href{https://www.tableau.com/pricing/teams-orgs}{Tableau} & 6 & 41 & 3 & 7 & 48 &
\href{https://www.dropbox.com/plans}{Dropbox} & 4 & 82 & 4 & 0 & 4 \\ \hline
\href{https://zapier.com/pricing}{Zapier} & 5 & 51 & 4 & 4 & 40 &
\href{https://evernote.com/compare-plans}{Evernote} & 6 & 32 & 4 & 0 & 4 \\ \hline
\rowcolor{gray!20}
\href{https://slack.com/pricing}{Slack} & 4 & 44 & 4 & 4 & 21 &
\href{https://hypercontext.com/pricing}{Hypercontext} & 4 & 63 & 4 & 0 & 4 \\ \hline
\href{https://mailchimp.com/es/pricing/marketing/compare-plans/?currency=USD}{MailChimp} & 6 & 90 & 4 & 5 & 15 &
\href{https://pumble.com/pricing}{Pumble} & 4 & 34 & 4 & 0 & 4 \\ \hline
\rowcolor{gray!20}
\href{https://clickup.com/pricing}{ClickUp} & 6 & 135 & 4 & 2 & 13 &
\href{https://userguiding.com/pricing}{UserGuiding} & 5 & 59 & 3 & 1 & 4 \\ \hline
\href{https://planable.io/pricing/}{Planable} & 6 & 41 & 4 & 2 & 13 &
\href{https://www.crowdcast.io/pricing}{Crowdcast} & 5 & 16 & 3 & 0 & 3 \\ \hline
\rowcolor{gray!20}
\href{https://clockify.me/pricing}{Clockify} & 6 & 72 & 6 & 4 & 10 &
\href{https://www.deskera.com/pricing}{Deskera} & 4 & 100 & 3 & 0 & 3 \\ \hline

\href{https://www.microsoft.com/en-us/microsoft-365/enterprise/office365-plans-and-pricing}{Microsoft 365} & 6 & 60 & 4 & 1 & 8 &
\href{https://www.overleaf.com/user/subscription/plans}{Overleaf} & 6 & 16 & 3 & 0 & 3 \\ \hline
\rowcolor{gray!20}
\href{https://trustmary.com/pricing/}{Trustmary} & 5 & 45 & 4 & 1 & 8 &
\href{https://quip.com/about/pricing}{Quip} & 6 & 15 & 3 & 0 & 3 \\ \hline
\end{tabularx}
\caption{Benchmark overview from \cite{ICSOC2024}. Each stat indicates the number of: snapshots (S), features (F), plans (P), add-ons (A), configuration space size (C); in 2024}
\label{table:dataset2024}
\vspace{-1cm}
\end{table}

The second dataset contains synthetic pricings intentionally designed to include inconsistencies. Although most cases adhere to the syntax rules of Pricing2Yaml, they exhibit errors that rendered them invalid within the iPricings domain. Thus, inconsistencies range from syntax issues (e.g., add-ons linked to nonexistent plans) to structural flaws, where pricings failed to behave as expected. Structural flaws include cases where the expected configuration space is not generated or “dead” pricing elements are present ---e.g., plans with identical features and usage limits but different prices. By applying the “valid subscription” operation across all cases, our framework successfully identified the majority of inconsistencies, while cases involving unaddressed analysis operations highlight the need for extending the approach to handle more complex scenarios like dead elements,\footnote{For further details, we invite the reader to consult the technical report \cite{LABPACK}.} paving the way for future advancements in pricing analysis.

As a final test, we applied our approach to replicate the analytical operations described in \cite{ICSOC2024} using their proposed benchmark. The results obtained with our method matched the manually extracted values reported in \cite{ICSOC2024}, demonstrating the accuracy and applicability of the proposed catalog of operations. 

In summary, the proposed framework demonstrates its robustness and versatility in automating iPricing analysis, addressing both real-world scenarios and deliberately challenging synthetic cases. By leveraging the modularity and scalability of MiniZinc, coupled with the support of tools like the Pricing2Yaml editor, the approach not only helps to reduce human errors but also provides an efficient pathway for analyzing pricings. The results highlight the potential of constraint programming to drive innovation in this domain, setting the foundation for future enhancements and broader applications in dynamic pricing~ecosystems.

\section{Conclusions and Future Work}
\label{sec:conclusions}

In this work, we first formalized iPricings as Constraint Satisfaction Optimization Problems (CSOPs), establishing a structured foundation for their analysis. Building on this, we defined a set of analysis operations to partially automate and streamline key pricing management tasks. For instance, evaluating the cardinality of the configuration space can aid decision-making during pricing design, while optimizing subscriptions based on cost or feature requirements can assist users in selecting the most suitable subscription.

As part of our testing efforts, we i) identified and corrected inconsistencies in a real-world SaaS pricing dataset \cite{ICSOC2024}, enhancing its reliability for future studies; and ii) developed a synthetic dataset to benchmark the detection of pricing inconsistencies, offering a robust foundation for evaluating and improving future contributions in this domain; iii) introduced a Pricing2Yaml editor within SPHERE, designed to improve the transition from web pricings to iPricings by reducing human error through real-time error detection and live rendering of pricings during their construction; and iv) also integrated within this platform a solution based on our approach: pricing cards\footnote{Sample pricing card can be found \href{https://sphere.score.us.es/pricings/sphere/Buffer?collectionName=CAISE\%202025}{here}}.

Looking ahead, this work opens several promising avenues for future research in automated analysis of SaaS pricings. Key opportunities include:

\begin{enumerate}
    \item Improving the explainability of the framework’s insights for developers and pricing decision-makers is critical. Given MiniZinc’s inherent limitations in providing explanations, exploring alternative solvers with built-in diagnostic capabilities could significantly enhance usability. However, such alternatives would require additional implementation effort to replicate the modularity that MiniZinc currently offers natively.

    \item Since our proposed framework is inherently tied to the Pricing2Yaml serialization—and thus, indirectly, to the Pricing4SaaS metamodel—it naturally inherits the limitations identified in prior research \cite{ICSOC2024}, such as the representation of bundles and discount mechanisms. A promising future direction is, therefore, to first address and extend these identified limitations in the Pricing4SaaS metamodel itself, before incorporating the corresponding improvements into our framework. This would allow the framework to accurately capture more intricate and realistic pricing strategies encountered in real-world SaaS offerings.

    \item The framework currently assumes that users express their requirements using the provider’s terminology, which is rarely the case. Developing an automated mapping process capable of translating user-defined requirements into the provider’s pricing terminology would significantly enhance the intuitiveness and practical applicability of the optimization operation of the framework.

    \item Manual translation from user-oriented pricings to iPricings introduces significant error risk, as highlighted by our results (see Section \ref{sec:tooling-support-and-validation}). Therefore, further automation in this process, possibly leveraging large language models (LLMs), shows promise for significantly reducing human errors.
\end{enumerate}

By addressing these challenges, we hope that future research can push the boundaries of Pricing-driven DevOps, making automated pricing analysis not only scalable and efficient but also intuitive, adaptable, and truly aligned with the evolving needs of SaaS ecosystems.

\begin{credits}
\subsubsection{\ackname} This work has been partially supported by grants PID2021-126227NB-C21 and PID2021-126227NB-C22 funded by MCIN / AEI / 10.13039 / 501100011033/FEDER, UE, and grants TED2021-131023B-C21 and TED2021-131023B-C22 funded by MCIN/AEI/10.13039/501100011033 and by European Union  “NextGenerationEU”/PRTR.

\end{credits}

\bibliographystyle{splncs04}
\bibliography{bibliography}

\begin{thebibliography}{10}
\providecommand{\url}[1]{\texttt{#1}}
\providecommand{\urlprefix}{URL }
\providecommand{\doi}[1]{https://doi.org/#1}

\bibitem{Caise2005}
Benavides, D., Trinidad, P., Ruiz-Cort{\'e}s, A.: Automated reasoning on feature models. In: International Conference on Advanced Information Systems Engineering. pp. 491--503. Springer (2005)

\bibitem{benavides10}
{Benavides, D., Segura, S., Ruiz-Cortés, A.}: {{Automated Analysis of Feature Models 20 Years Later: A Literature Review}}. {Information Systems}  \textbf{{35}}({6}),  {615 -- 636} ({2010})

\bibitem{AI4PRICING}
Cavero, F.J., Alonso, J.C., Ruiz-Cortés, A.: {From Static to Intelligent: Evolving SaaS Pricing with LLMs}. In: Proceeding of the International Conference on Service Oriented Computing, (ICSOC). 1st International Workshop on Service-Oriented Computing for Artificial Intelligence, (SOC4AI), Springer LNCS (2024)

\bibitem{FOWLER2023}
Fowler, M.: Feature toggles (aka feature flags) (nd), \url{https://martinfowler.com/articles/feature-toggles.html}, accessed: December 2023

\bibitem{garcia2021flexible}
Garc{\'\i}a, J.M., Mart{\'\i}n-D{\'\i}az, O., Fernandez, P., M{\"u}ller, C., Ruiz-Cort{\'e}s, A.: A flexible billing life cycle for cloud services using augmented customer agreements. IEEE Access  \textbf{9},  44374--44389 (2021)

\bibitem{ICSOC2024}
Garc{\'i}a-Fern{\'a}ndez, A., Parejo, J.A., Cavero, F.J., Ruiz-Cort{\'e}s, A.: Racing the market: An industry support analysis for pricing-driven devops in saas. In: Service-Oriented Computing. pp. 260--275 (2025)

\bibitem{CAISEFORUM2024}
Garc{\'\i}a-Fern{\'a}ndez, A., Parejo, J.A., Ruiz-Cort{\'e}s, A.: {Pricing4SaaS}: Towards a pricing model to drive the operation of {SaaS}. In: Intelligent Information Systems, - CAiSE Forum, Proceedings. Lecture Notes in Business Information Processing, vol.~520, pp. 47--54. Springer (2024)

\bibitem{ICWE2024}
Garc{\'i}a-Fern{\'a}ndez, A., Parejo, J.A., Trinidad, P., Ruiz-Cort{\'e}s, A.: {Towards Pricing4SaaS: A Framework for Pricing-Driven Feature Toggling in SaaS}. In: Web Engineering. 24th International Conference, {ICWE}. pp. 389--392. Springer (2024)

\bibitem{LABPACK}
García-Fernández, A., Parejo, J.A., Trinidad, P., Ruiz-Cortés, A.: {Automated Analysis of Intelligent Pricings - Supplementary Material} (2024). \doi{10.5281/zenodo.14254341}

\bibitem{Jiang2009}
Jiang, Z., Sun, W., Tang, K., Snowdon, J., Zhang, X.: A pattern-based design approach for subscription management of software as a service. 2009 Congress on Services - I pp. 678--685 (2009)

\bibitem{Muller18TSC}
M{\"u}ller, C., Gutierrez~Fernandez, A.M., Fernandez, P., Martin-Diaz, O., Resinas, M., Ruiz-Cortes, A.: {{A}utomated {V}alidation of {C}ompensable {SLA}s}. IEEE Transactions on Services Computing  (2018)

\bibitem{Muller13TSC}
M{\"u}ller, C., Resinas, M., Ruiz-Cortés, A.: {{A}utomated {A}nalysis of {C}onflicts in {WS--A}greement}. IEEE Transactions on Services Computing  \textbf{7}(4),  530--544 (2014)

\bibitem{muller2013automated}
M{\"u}ller~Cej{\'a}s, C., Resinas Arias~de Reyna, M., Ruiz~Cort{\'e}s, A.: Automated analysis of conflicts in ws-agreement documents. IEEE Transactions On Services Computing, 7 (4), 530-544  (2013)

\bibitem{minizinc}
Nethercote, N., Stuckey, P.J., Becket, R., Brand, S., Duck, G.J., Tack, G.: Minizinc: Towards a standard cp modelling language. In: International Conference on Principles and Practice of Constraint Programming. pp. 529--543. Springer (2007)

\bibitem{Hofstede98}
Ter~Hofstede, A.H.M., Proper, H.A.: {{H}ow to {F}ormalize {I}t? {F}ormalization {P}rinciples for {I}nformation {S}ystem {D}evelopment {M}ethods}. Information and Software Technology  \textbf{40},  519--540 (1998)

\bibitem{Trinidad2013}
Trinidad, P., Ruiz-Cort{\'e}s, A., Benavides, D.: Automated analysis of stateful feature models. In: Seminal Contributions to Information Systems Engineering: 25 Years of CAiSE. pp. 375--380 (2013). \doi{10.1007/978-3-642-36926-1_30}

\end{thebibliography}

%
%
%
%




\end{document}